\documentclass[pra,twocolumn,twoside,showpacs]{revtex4}

\usepackage{graphicx,dcolumn,color}
\usepackage{amsmath,amssymb,bm}

\newcommand{\eq}[1]{\begin{equation}#1\end{equation}}
\newcommand{\eqmulti}[1]{\begin{equation}\begin{split}#1\end{split}\end{equation}}

\newcommand{\bra}[1]{\ensuremath{ \big< {#1} \big| \, }}
\newcommand{\ket}[1]{\ensuremath{ \, \big| {#1} \big> }}
\newcommand{\braket}[2]{\ensuremath{ \big< {#1} \big| {#2} \big> }}

\newcommand{\matrixe}[3]{\ensuremath{ \big< {#1} \big| \,{#2}\, \big| {#3} \big> }}

\newcommand{\op}[1]{\ensuremath{\bm{#1}}}
\newcommand{\corr}[1]{\ensuremath{\tilde{#1}}}
\newcommand{\adj}[1]{\ensuremath{{{#1}}^{\dag}}}

\newcommand{\ii}{\ensuremath{\mathrm{i}}}
\newcommand{\dd}{\ensuremath{\mathrm{d}}}
\newcommand{\partd}[2]{\ensuremath{\frac{\partial{#1}}{\partial{#2}}}}
\newcommand{\partdd}[2]{\ensuremath{\frac{\partial^2{#1}}{\partial{#2}^2}}}

\newcommand{\gO}{\ensuremath{\op{g}}}

\newcommand{\rO}{\ensuremath{\op{r}}}
\newcommand{\tO}{\ensuremath{\op{t}}}
\newcommand{\uO}{\ensuremath{\op{u}}}
\newcommand{\vO}{\ensuremath{\op{v}}}
\newcommand{\wO}{\ensuremath{\op{w}}}

\newcommand{\CO}{\ensuremath{\op{C}}}
\newcommand{\CCO}{\ensuremath{\adj{\op{C}}}}
\newcommand{\GO}{\ensuremath{\op{G}}}
\newcommand{\HO}{\ensuremath{\op{H}}}

\newcommand{\OO}{\ensuremath{\op{O}}}
\newcommand{\PO}{\ensuremath{\op{P}}}
\newcommand{\RO}{\ensuremath{\op{R}}}
\newcommand{\TO}{\ensuremath{\op{T}}}
\newcommand{\UO}{\ensuremath{\op{U}}}
\newcommand{\VO}{\ensuremath{\op{V}}}

\newcommand{\XO}{\ensuremath{\op{X}}}

\newcommand{\idO}{\ensuremath{\op{1}}}

\newcommand{\pV}{\ensuremath{\vec{p}}}

\newcommand{\rV}{\ensuremath{\vec{r}}}
\newcommand{\xV}{\ensuremath{\vec{x}}}

\newcommand{\XV}{\ensuremath{\vec{X}}}

\newcommand{\rN}{\ensuremath{\hat{r}}}

\newcommand{\lOV}{\ensuremath{\vec{\op{l}}}}
\newcommand{\pOV}{\ensuremath{\vec{\op{p}}}}
\newcommand{\qOV}{\ensuremath{\vec{\op{q}}}}
\newcommand{\rOV}{\ensuremath{\vec{\op{r}}}}
\newcommand{\xOV}{\ensuremath{\vec{\op{x}}}}

\newcommand{\POV}{\ensuremath{\vec{\op{P}}}}
\newcommand{\XOV}{\ensuremath{\vec{\op{X}}}}

\newcommand{\rON}{\ensuremath{\hat{\op{r}}}}

\newcommand{\Rm}{\ensuremath{R_-}}

\newcommand{\RRm}{\ensuremath{\mathcal{R}_-}}

\newcommand{\Rp}{\ensuremath{R_+}}
\newcommand{\DRp}{\ensuremath{R'_+}}
\newcommand{\DDRp}{\ensuremath{R''_+}}
\newcommand{\DDDRp}{\ensuremath{R'''_+}}

\newcommand{\RRp}{\ensuremath{\mathcal{R}_+}}

\newcommand{\Rpm}{\ensuremath{R_{\pm}}}
\newcommand{\Rmp}{\ensuremath{R_{\mp}}}

\newcommand{\RRpm}{\ensuremath{\mathcal{R}_{\pm}}}

\newcommand{\DeltapmV}{\ensuremath{\vec{\Delta}_{\pm}}}
\newcommand{\DeltapV}{\ensuremath{\vec{\Delta}_{+}}}

\newcommand{\Xp}{\ensuremath{\mathcal{X}_+}}
\newcommand{\Xm}{\ensuremath{\mathcal{X}_-}}
\newcommand{\Xpm}{\ensuremath{\mathcal{X}_{\pm}}}

\newcommand{\Dm}{\ensuremath{\mathcal{D}_-}}

\newcommand{\cm}{\ensuremath{\mathrm{cm}}}
\newcommand{\equi}{\ensuremath{\mathrm{eq}}}
\newcommand{\ex}{\ensuremath{\mathrm{ex}}}
\newcommand{\intr}{\ensuremath{\mathrm{int}}}
\newcommand{\rel}{\ensuremath{\mathrm{rel}}}
\newcommand{\sat}{\ensuremath{\mathrm{sat}}}

\begin{document}

\title{Short-Range Correlations in ${}^4$He Liquid and Small 
  ${}^4$He Droplets\\ described by the Unitary Correlation Operator Method }

\author{R. Roth}
\email{r.roth@gsi.de}

\author{H. Feldmeier}
\email{h.feldmeier@gsi.de}

\affiliation{Gesellschaft f\"ur Schwerionenforschung (GSI), 
  Planckstr. 1, 64291 Darmstadt, Germany}

\date{\today}

\begin{abstract}
\vskip5pt 

The Unitary Correlation Operator Method (UCOM) is employed to treat
short-range correlations in both, homogeneous liquid and small
droplets of bosonic $^4$He atoms. The dominating short-range
correlations in these systems are described by an unitary
transformation in the two-body relative coordinate, applied either to
the many-body state or to the Hamiltonian and other operators.  It is
shown that the two-body correlated interaction can describe the
binding energy of clusters of up to 6 atoms very well, the numerical
effort consisting only in calculating one two-body matrix element with
Gaussian single-particle states. The increasing density of bigger
droplets requires the inclusion of correlation effects beyond the
two-body order, which are successfully implemented by a
density-dependent two-body correlator. With only one adjusted
parameter the binding energies and radii of larger droplets and the
equation of state of the homogeneous $^4$He liquid can be described
quantitatively in a physically intuitive and numerically simple way.

\end{abstract}

\pacs{67.40.Db 61.20.Gy 36.40.-c}
%

\maketitle



\section{Introduction}

The quantum-mechanical description of ground state properties of
interacting many-body systems is a long-standing challenge in
theoretical physics. In general the problem is two-fold: The first
task is to specify the interaction between the constituents in terms
of a two-body potential. The second is to solve the quantum-mechanical
many-body problem with that interaction.

For many-body systems composed of Helium atoms the first task, the
construction of an appropriate interatomic potential, is solved rather
precisely. The HFDHE2 potential of Aziz \emph{et al.} \cite{AzNa79} is
able to reproduce experimental transport coefficients in a wide
temperature range. A simpler and less precise description of the
interaction is given by the Lennard-Jones potential \cite{BoMi38,
BeKe81}
\eq{ \label{eq:intro_lennardjones}
 v(r) = 4\epsilon\; [(\sigma/r)^{12} - (\sigma/r)^6]
}
with $\epsilon = 10.22\text{K}$ and $\sigma = 2.556\text{\AA}$.  Since
the emphasis of the investigations presented in this paper is more on
novel concepts, which remarkably simplify the treatment of the
many-body problem, we will restrict ourselves to this simple and
widely used parameterization.

A common property of many realistic microscopic interactions --- like
van der Waals type potentials between neutral atoms or the
nucleon-nucleon interaction in atomic nuclei --- is a strong
short-range repulsion, which induces strong short-range correlations
in the quantum many-body state. The repulsive core prevents the
particles to approach each other closer than the radius of this core
such that the two-body density matrix is strongly depleted for
particle distances smaller than the core radius. This effect cannot be
described in a mean-field like picture or in terms of a superposition
of a finite number of symmetrized product states.

In this work we treat these correlations with the Unitary Correlation
Operator Method (UCOM), that has already been applied successfully in
the framework of the nuclear many-body problem \cite{UCOM98}. 

In the following section we introduce the general formalism of the
Unitary Correlation Operator Method and derive analytic expressions
for correlated wave functions and the correlated Hamiltonian. In the
third section we determine the optimal correlation operator for liquid
$^4$He from the two-body system and discuss the structure of the
effective interaction generated by the unitary transformation. In
section IV we investigate the homogeneous $^4$He liquid with that
correlator and introduce a density-dependent two-body correlator that
describes very well the influence of three- and more-body
correlations. Finally, we successfully apply the density-dependent
correlator to describe the ground state of small $^4$He
droplets. Considering the great numerical simplification --- only a
few one-dimensional integrals have to be calculated --- compared to
Monte-Carlo-type calculations \cite{KaLe81,ScSc65}, the agreement of
the results is excellent.

\clearpage
\section{Unitary Correlation Operator Method (UCOM)}
\label{sec:ucom}

\subsection{Concept}
\label{sec:ucom_concept}

The goal of the Unitary Correlation Operator Method (UCOM) is to
describe short-range interaction-induced correlations by a unitary
transformation of an uncorrelated many-body state with $N$
particles. The unitary operator $\CO$ associated with this
transformation can be written as an exponential of a hermitian
generator $\GO$
\eq{ \label{eq:ucom_correlator}
  \CO = \exp[-\ii\, \GO] .
}
The generator is chosen to be an irreducible two-body operator
\eq{ \label{eq:ucom_generator}
  \GO = \sum_{i<j}^N \gO_{ij} ,
}
which describes genuine two-body correlations only. In principle we
could add a three-body part to account for genuine three-body
correlations. For the moment we omit this step to develop the basic
formalism on a simple level.

The particular structure of the unitary transformation --- which
should reflect the properties of the short-range correlations --- is
described by the two-body generator $\gO$. The repulsive core of the
interaction leads to a suppression of the probability density within
the range $r_{\text{core}}$ of the core. Uncorrelated many-body states
that are symmetrized products of single-particle states lead, however,
to a two-body density that does not show any depletion for relative
distances $r<r_{\text{core}}$ between the pair of particles. Therefore
the unitary transformation should shift those particles that are in the
forbidden region of the repulsion to larger distances. This is done by
means of a distance-dependent radial shift in the relative coordinate
of two-particles. The hermitian generator of this shift in two-body
space reads
\eq{ \label{eq:ucom_generator_2b}
  \gO = \frac{1}{2}\big[ s(\rO)\; (\rON\cdot\qOV) 
    + (\qOV\cdot\rON)\; s(\rO) \big] ,
}
where $\qOV = \frac{1}{2}(\pOV_1 - \pOV_2)$ is the operator of the
relative momentum, $\rOV=\xOV_1-\xOV_2$ is the relative coordinate, and
$\rON$ is the associated unit vector. The function $s(r)$ describes
the size of the shift as function of the distance of the particles.

\subsection{Correlated Wave Functions}
\label{sec:ucom_wavefct}

First we want to apply the unitary correlation operator $\CO$ to an
uncorrelated many-body state $\ket{\Psi}$ out of a low-momentum
model space. The correlated many-body state $\ket{\corr{\Psi}}$ is
defined by
\eq{ \label{eq:ucom_corr_state}
  \ket{\corr{\Psi}} = \CO \ket{\Psi} .
}
Here and in the following all correlated quantities are marked by a
tilde.

In order to demonstrate the generic effect of the correlation operator we
discuss first a two-body system with an uncorrelated two-body state
$\ket{\psi}$. Using the definition of the correlation operator $\CO$
and the generator $\gO$ \eqref{eq:ucom_generator_2b} we can evaluate
the correlated two-body wave function in coordinate representation
explicitly \cite{UCOM98}
\eq{ \label{eq:ucom_corr_wavefct_1}
 \matrixe{\XV,\rV}{\CO\,}{\psi}
  = \RRm(r)\; \braket{\XV,\Rm(r)\, \rN}{\psi} . 
}
The respective transformation with the hermitian adjoint correlation
operator $\CCO$ leads to
\eq{ \label{eq:ucom_corr_wavefct_2}
  \matrixe{\XV,\rV}{\adj{\CO}\!}{\psi}
  = \RRp(r)\; \braket{\XV,\Rp(r)\, \rN}{\psi} .
}
The correlation operator acts in the relative coordinate $\rV$ of the
two-body system only, the center of mass coordinate
$\XV=\frac{1}{2}(\xV_1+\xV_2)$ is invariant. According to the
construction of the the generator \eqref{eq:ucom_generator_2b} the
unitary transformation corresponds to a radial shift in the relative
two-body coordinate
\eq{ \label{eq:ucom_coordtrafo}
  \rV\; \overset{\CO}{\longmapsto}\; \Rm(r)\,\rN 
  \;,\qquad
  \rV\; \overset{\CCO}{\longmapsto}\; \Rp(r)\,\rN .
}
The transformed relative distances are given by the correlation
functions $\Rm(r)$ and $\Rp(r)$, respectively.  Due the unitarity of the
correlation operator $\CO\CCO=\idO$ the correlation functions $\Rm(r)$
and $\Rp(r)$ are inverse to each other
\eq{ \label{eq:ucom_corr_corrfunc_inverse}
  \Rpm[\Rmp(r)] = r .
}
The conservation of the norm of the correlated two-body wave function
is guaranteed by the metric factor $\RRpm(r)$ in
\eqref{eq:ucom_corr_wavefct_1} and \eqref{eq:ucom_corr_wavefct_2}
\eq{ \label{eq:ucom_corr_metric}
  \RRpm(r) = \frac{\Rpm(r)}{r} \sqrt{\Rpm'(r)} ,
}
where $\Rpm'(r)$ denotes the derivative of $\Rpm(r)$. It corresponds
to the square root of the Jacobi determinant associated with the
transformation of the normalization integral.

The correlation functions $\Rpm(r)$ that describe the coordinate
transformation are connected to the function $s(r)$, which enters in
the generator \eqref{eq:ucom_generator_2b}, by the integral equation
\eq{
  \pm1 = \int_r^{\Rpm(r)}\!\! \frac{\dd\xi}{s(\xi)} .
}
For small shift distances the approximate relation
\eq{
  \Rpm(r) \approx r \pm s(r)
}
holds.

Since all correlated quantities can be expressed directly in terms of
the correlation functions $\Rm(r)$ and $\Rp(r)$ we will use one of
these as basic quantity to describe the detailed structure of the
short-range correlations. The shift function $s(r)$ is only used for
formal considerations and never specified explicitly.

\subsection{Correlated Operators}
\label{sec:ucom_operator}

Complementary to the definition of a correlated state
\eqref{eq:ucom_corr_state} we can use the correlation operator to
define correlated operators. For an arbitrary observable $\OO$ the
correlated operator $\corr{\OO}$ is given by the similarity
transformation
\eq{
  \corr{\OO} = \CCO \OO\, \CO .
}
It is evident that the formulations in terms of correlated states and
correlated operators are equivalent when expectation values or
matrix elements are calculated
\eq{
  \matrixe{\corr{\Psi}}{\OO}{\corr{\Psi}'} 
  = \matrixe{\Psi}{\CCO\OO\,\CO}{\Psi'}
  = \matrixe{\Psi}{\corr{\OO}}{\Psi'} .
}
Thus we can choose the formulation that is technically or intuitively
better suited for the specific application under consideration.  In
most cases the formulation in terms of correlated operators is easier
to apply in the many-body system. Moreover it provides a systematic
approximation scheme that will be discussed in section
\ref{sec:ucom_clusterexp}. 

As for the correlated wave function we first discuss specific
correlated operators for a two-body system. The generalization to
many-body systems will be done in the following sections.

\subsubsection{Correlated Two-Body Potential}

The simplest operator of interest is the local two-body potential
$v(\rO)$ that depends on the relative distance $\rO$ only. To evaluate
the correlated potential we use the results on the correlated two-body
wave function to determine a general two-body matrix element of the form
\eqmulti{
  &\matrixe{\psi}{\CCO v(\rO)\CO}{\psi'} \\[1pt]
  &\quad= \int\!\!\dd^3r\,\dd^3\!X\;\matrixe{\psi}{\CCO}{\XV,\rV}
    \,v(r)\, \matrixe{\XV,\rV}{\CO}{\psi'} \\[-2pt] 
  &\quad= \int\!\!\dd^3r\,\dd^3\!X\;\braket{\psi}{\XV,\rV}
    \;v[\Rp(r)]\; \braket{\XV,\rV}{\psi'} \\[2pt]
  &\quad= \matrixe{\psi}{v[\Rp(\rO)]}{\psi'} .
}
In going from the second to the third line we use equations
\eqref{eq:ucom_corr_wavefct_1} and
\eqref{eq:ucom_corr_corrfunc_inverse} and substitute the integration
variable $\rV$ by $\Rm(r)\,\rN$. Since the two-body states
are arbitrary this relation is valid on the operator level
\eq{ \label{eq:ucom_corr_pot}
  \corr{v}(\rO) 
  = \CCO v(\rO) \CO
  = v[\Rp(\rO)] .
}
Thus the correlated two-body potential $\corr{v}(r)$ is given by the
original potential with transformed radial coordinate $r\mapsto\Rp(r)$.

\subsubsection{Correlated Kinetic Energy}

The correlated operator of the kinetic energy in the two-body system
is slightly more complicated due to its momentum dependence. First we
decompose the operator of the two-body kinetic energy $\TO_2$ in a
center of mass contribution $\tO_{\cm}$ and an relative part
$\tO_{\rel}$
\eq{
  \TO_2 = \tO_{\cm} + \tO_{\rel}
  = \frac{1}{2M} \POV^2 + \frac{1}{2\mu} \qOV^2 .
}
The operator of the center of mass momentum $\POV=\pOV_1+\pOV_2$
commutes with the correlation operator, i.e., the operator of the
center of mass kinetic energy is invariant under the unitary
transformation. The correlation operator acts on the relative part
$\tO_{\rel}$ of the kinetic energy only. A similar calculation as for
the local potential leads to the following structure of the correlated
relative kinetic energy \cite{UCOM98}
\eq{ \label{eq:ucom_corr_kin}
  \corr{\tO}_{\rel} = \CCO\tO_{\rel}\,\CO 
  = \tO_{\rel} + \corr{\tO}_{\Omega} + \corr{\tO}_{r} + \corr{u}(\rO) .
}
Besides $\tO_{\rel}$ an additional momentum dependent term appears in
the correlated relative kinetic energy, which can be formulated in
terms of a tensorial effective mass correction.  It is conveniently
split into a correction for the radial part $\corr{\tO}_r$ and a
different correction for the angular component $\corr{\tO}_{\Omega}$
\eq{ \label{eq:ucom_corr_effmasscorr}
  \corr{\tO}_{\Omega}
  = \frac{1}{2\corr{\mu}_{\Omega}(\rO)}\, \frac{\lOV^2}{\rO^2}
  \;,\qquad
  \corr{\tO}_{r} 
  = (\qOV\cdot\rON) \frac{1}{2\corr{\mu}_r(\rO)} (\rON\cdot\qOV) ,
}
where $\lOV=\rOV\times\qOV$ is the operator of the relative two-body
angular momentum.  The distance dependent effective mass corrections
$\corr{\mu}_{\Omega}(\rO)$ and $\corr{\mu}_r(\rO)$ depend on the
correlation function $\Rp(r)$ only
\eq{ \label{eq:ucom_corr_effmass}
  \frac{\mu}{\corr{\mu}_{\Omega}(r)} = \frac{r^2}{\Rp^2(r)} - 1
  \;,\qquad
  \frac{\mu}{\corr{\mu}_r(r)} = \frac{1}{\DRp{}^{\!\!2}(r)} - 1 .
}
In addition to these momentum dependent terms a local contribution
$\corr{u}(\rO)$ appears in \eqref{eq:ucom_corr_kin}
\eqmulti{ \label{eq:ucom_corr_kinpot}
\corr{u}(r) &= \frac{1}{2\mu} \frac{1}{\DRp{}^{\!\!2}(r)} \\ 
  &\;\;\times\bigg[ 2\frac{\DDRp(r)}{r\,\DRp(r)} 
  - \frac{5}{4}\bigg(\frac{\DDRp(r)}{\DRp(r)}\bigg)^{\!\!2}
  + \frac{1}{2}\,\frac{\DDDRp(r)}{\DRp(r)} \bigg] .
}
This additional two-body potential is also determined by the
correlation function $\Rp(r)$ and its derivatives.

\subsection{Cluster Expansion and Two-Body Approximation}
\label{sec:ucom_clusterexp}

In the preceding section we evaluated correlated operators explicitly
in the two-body system. However, correlated operators in the many-body
system contain additional terms.

The unitary transformation of an operator, e.g., an one-body operator
like the kinetic energy or a two-body operator like the potential,
generates a correlated operator, which contains irreducible
contributions to all particle numbers
\eqmulti{ \label{eq:ucom_clusterexp} 
  \corr{\OO} 
  &= \CCO \OO\, \CO \\ 
  &= \corr{\OO}^{[1]} + \corr{\OO}^{[2]} + \corr{\OO}^{[3]} + \dots .
}
We use the notation $\corr{\OO}^{[k]}$ for the irreducible
$k$-body part of the correlated operator $\corr{\OO}$. This
decomposition according to irreducible particle number is called
cluster expansion \cite{UCOM98,Clar79}.

The cluster expansion gains physical meaning by the so called cluster
decomposition principle \cite{Clar79}: Two localized subsystems, which
are separated beyond the range of interactions, are independent of
each other. Therefore the state of the total system decouples into a
direct product of the states of the two subsystems. This implies an
analogous decomposition property of the correlation operator and of
correlated operators.

Because of the cluster decomposition principle the cluster expansion
is a natural starting point for approximations of the correlated
operators \eqref{eq:ucom_clusterexp}. For a selected particle there
will be a certain number of other particles within the range of the
correlation depending on the density. The number of particles in this
cluster gives the maximum order of the cluster expansion that
contributes. We can truncate the cluster expansion at low orders if
the range of the correlations is sufficiently small compared to the
average distance of the particles.

The simplest nontrivial approximation results from the truncation of
the cluster expansion beyond two-body order. The correlated operator
in two-body approximation reads
\eq{
  \corr{\OO}^{C2} = \corr{\OO}^{[1]} + \corr{\OO}^{[2]} .
}
This approximation requires that the system is sufficiently dilute
such that contributions of the three-body order of the cluster
expansion are small. Or in other words, the probability for three and
more particles to be in the range of the repulsive core simultaneously
has to be small. The technical advantage of the two-body
approximation is that closed analytic expressions for the correlated
operators can be deduced by just considering the two-body system. The
correlated Hamiltonian in two-body approximation is of the form
\eqmulti{
  \corr{\HO}^{C2} 
  &= \corr{\TO}^{[1]} + \corr{\TO}^{[2]} + \corr{\VO}^{[2]} \\
  &= \sum_i^N \corr{\tO}^{[1]}_i + \sum_{i<j}^N \corr{\tO}^{[2]}_{ij}
    + \sum_{i<j}^N \corr{\vO}^{[2]}_{ij} .
}
The one-body part of the correlated kinetic energy is simply the
uncorrelated kinetic energy operator. The two-body part consists of
the effective mass corrections \eqref{eq:ucom_corr_effmasscorr} and the
additional local potential \eqref{eq:ucom_corr_kinpot} as determined
in the previous section 
\eqmulti{
  \corr{\tO}^{[1]} &= \tO \\
  \corr{\tO}^{[2]} &= \corr{\tO}_{\Omega} + \corr{\tO}_r
    + \corr{u}(\rO) .   
}
The correlated potential has only a two-body contribution that is
given by \eqref{eq:ucom_corr_pot}
\eq{ \label{eq:ucom_corr_pot2}
  \corr{\vO}^{[2]} = \corr{v}(\rO) .
}

In order to get a rough measure for the validity of the two-body
approximation we define a smallness parameter
\eq{ \label{eq:ucom_smallnesspara}
  \kappa = \rho\, V_C
}
as a product of the density $\rho$ of the system and a typical volume
in which the correlations between a pair of particles change their
relative wave function. The correlation volume $V_C$ is defined by the
norm of the defect wave function, i.e., the difference between the
uncorrelated uniform wave function $\braket{\rV}{\phi_0}=1$ and its
correlated companion
\eq{ \label{eq:ucom_corrvol}
  V_C 
  = \int\!\!\dd^3r\; [\RRm(r)-1]^2 
  = \int\!\!\dd^3r\; [\RRp(r)-1]^2 .
}
The smallness parameter $\kappa$ is a measure for the probability to
find a third particle within the volume where the correlations between
two particles change their relative wave function significantly. The
two-body approximation is valid only if this probability is small such
that three-body correlations are negligible. We have shown for
different physical systems \cite{Roth00,UCOM98} that the relative
contribution of the three-body order to the energy exceeds 10\% if the
smallness parameter reaches a value of typically $\kappa=0.3$.

\subsection{Many-Body Correlations}
\label{sec:ucom_mb_corr}

If the two-body approximation is not sufficient one can include higher
orders of the cluster expansion successively. However, already the
calculation of a three-body correlated wave function starting from the
many-body correlation operator \eqref{eq:ucom_correlator} in analogy
to section \ref{sec:ucom_wavefct} is not practicable. Therefore we
reverse the procedure and start from a general many-body coordinate
transformation --- in analogy to the correlation function $\Rpm(r)$
--- and determine correlated many-body wave functions and
operators. In a second step we connect the explicit structure of the
coordinate transformation with the many-body correlation operator, at
least in an approximate way.

\subsubsection{Many-Body Coordinate Transformation}

Consider a $N$-body system with a collective coordinate vector
$X=(\xV_1,\xV_2,\dots,\xV_N)$. The short-range 
correlations are described by a $N$-body coordinate transformation
\eq{ \label{eq:ucom_mb_ctrafo}
  X \mapsto X' = \Xm(X) .
}
The transformation function $\Xm(X)$ corresponds to the correlation
function $\Rm(r)$ in the two-body case. Similar to $\Rp(r)$ we define
an inverse transformation function $\Xp(X)$ by
\eq{
  \Xp[\Xm(X)] = X . 
}
The correlated $N$-body wave function $\braket{X}{\corr{\Psi}}$ is
defined via the coordinate transformation \eqref{eq:ucom_mb_ctrafo}
\eq{
  \braket{X}{\corr{\Psi}} = \Dm(X)\; \braket{\Xm(X)}{\Psi}
}
with a metric factor $\Dm(X)$ that ensures norm conservation.  The
metric factor is given by the square root of the Jacobi determinant of
the transformation
\eq{
  \Dm(X) = || J_-(X) ||^{1/2}
}
with the matrix elements
\eq{
  [J_-(X)]_{ij} = \partd{X'_i}{X_j} 
    = \partd{[\Xm(X)]_i}{X_j} .
} 
Formally we can construct a unitary many-body correlation operator
$\CO$ that generates this particular coordinate transformation
\eq{ 
  \CO 
  = \int\!\! \dd^{3N}\!X \; \ket{X}\; \Dm(X)\; \bra{\Xm(X)} .
}
The check of the unitarity relation $\CCO\CO=\idO$ with this
definition is straightforward.

Using this formulation we can evaluate many-body correlated operators
in the same way as shown in section \ref{sec:ucom_operator} for the
two-body approximation. We will only present selected results that are
needed for the following investigations. The full formalism is
discussed in \cite{Roth00}.

A general local potential $V(X)$, which depends on the coordinates
$X=(\xV_1,\xV_2,\dots,\xV_N)$ of all $N$ particles, is transformed as
in the two-body case
\eq{ \label{eq:ucom_mb_corr_pot}
  \CCO V(\XO)\, \CO  = V[\Xp(\XO)] ,
}
i.e., the correlated potential is given by the original potential with
transformed coordinate dependence. 

The expression for the $N$-body correlated kinetic energy operator
$\CCO\TO\,\CO$ is more involved
\eq{
  \CCO \TO\, \CO
  = \sum_{j,k=1}^{3N} \PO_j \; [\corr{M}(\XO)]_{jk} \; \PO_k 
  + \corr{U}(\XO) ,
}
where $P=(\pV_1,\pV_2,\dots,\pV_N)$ denotes the collective momentum
vectors of all $N$ particles. As in the two-body case
\eqref{eq:ucom_corr_kin} the correlated kinetic energy contains an
effective mass tensor and an additional local potential. The effective
mass tensor is given by the square of the Jacobi matrix
\eq{
[\corr{M}(X)]_{jk} 
= \frac{1}{2m} \sum_{i=1}^{3N} [J_-(X)]_{ji}\, [J_-(X)]_{ik} .
}
The additional local potential $\corr{U}(X)$ given by
\eq{ \label{eq:ucom_mb_corr_kinpot}
  \corr{U}(X) 
  = -\frac{1}{2m} \bigg[ \frac{1}{\Dm(X')}\; \sum_{i=1}^{3N}
  \partdd{}{X'_i} \; \Dm(X') \bigg]_{X' = \Xp(X)} .
}
With these expressions it is in principle possible to calculate
expectation values of the correlated Hamiltonian up to arbitrary order
in the cluster expansion. In practice all extensions beyond two-body
approximation are very costly, thus only the three-body order
will be considered.

\subsubsection{Three-Body Correlations}

We will use the general formulation of the many-body coordinate
transformation to estimate the contribution of the three-body order of
the cluster expansion. Since we cannot derive the many-body coordinate
transformation directly from the correlation operator we will
construct the transformation by generalization of the known one in the
two-body system.

It is useful to reformulate the two-body coordinate transformation
discussed in section \ref{sec:ucom_wavefct} in terms of the general
transformation function $\Xpm^{(2)}(X)$, where the upper index
indicates the number of particles involved. The transformation
\eqref{eq:ucom_coordtrafo} of the relative coordinate with the
correlation function $\Rpm(r)$ is equivalent to the following
transformation of the single-particle coordinates
\eq{ \label{eq:ucom_ctrafo_2b}
\begin{array}{c @{\;\;} c @{\;\;} c @{\,} c @{\,} c}
\xV_1 &\overset{\Xpm^{(2)}}{\longmapsto}& \xV_1 &+& \DeltapmV(\rV_{12})
\\
\xV_2 &\overset{\Xpm^{(2)}}{\longmapsto}& \DeltapmV(\rV_{21}) &+& \xV_2
\end{array} 
}
with a shift vector
\eq{
\DeltapmV(\rV) = \tfrac{1}{2} [\Rpm(r)-r]\, \rN . 
}
This notation reveals the intuitive picture behind the description of
short-range correlations by means of a coordinate transformation. As a
consequence of the short-range repulsion the particles are displaced
along their connecting axis by the shift vector $\DeltapmV(\rV)$.

The two-body transformation \eqref{eq:ucom_ctrafo_2b} can be readily 
generalized to describe three-body correlations. The corresponding
three-body coordinate transformation $\Xp^{(3)}(X)$ is given by
\eq{ \label{eq:ucom_ctrafo_3b}
\begin{array}{c @{\;\;} c @{\;\;} c @{\,} c @{\,} c@{\,} c @{\,} c}
\xV_1 &\overset{\Xp^{(3)}}{\longmapsto}& \xV_1 &+&  
  \DeltapV(\rV_{12}) &+& \DeltapV(\rV_{13}) \\
\xV_2 &\overset{\Xp^{(3)}}{\longmapsto}& \DeltapV(\rV_{21}) &+&
  \xV_2 &+& \DeltapV(\rV_{23}) \\
\xV_3 &\overset{\Xp^{(3)}}{\longmapsto}& \DeltapV(\rV_{31}) &+& 
  \DeltapV(\rV_{32}) &+& \xV_3
\end{array} .
}
Thus the coordinate of the first particle is transformed with respect
to the second and the third particle simultaneously. As a necessary
prerequisite this transformation obeys the cluster decomposition
principle. If one of the three particles is separated beyond the range
of the correlations, i.e., to distances where $\DeltapmV(\rV)$
vanishes, then the transformation reduces to the two-body
transformation \eqref{eq:ucom_ctrafo_2b} for the remaining pair.

We will use this three-body transformation to calculate the local
three-body contributions of the correlated Hamiltonian, i.e., the
correlated potential and the local part of the correlated kinetic
energy. The irreducible three-body part $\corr{\vO}^{[3]}$ of the
correlated potential is given by
\eqmulti{
  \corr{\vO}^{[3]} 
  &= \CCO (\vO_{12} + \vO_{13} + \vO_{23})\, \CO \\  
  &- (\corr{\vO}^{[2]}_{12} + \corr{\vO}^{[2]}_{13} 
    + \corr{\vO}^{[2]}_{23}) ,
}
where $\vO_{ij}$ is the uncorrelated two-body potential and
$\corr{\vO}^{[2]}_{ij}$ is the two-body part of the correlated
potential. The first term describes the fully correlated two-body
potential in a three-body system, which is given by a three-body
coordinate transformation according to \eqref{eq:ucom_mb_corr_pot}.
The second term is the sum of the two-body correlated potentials
\eqref{eq:ucom_corr_pot2}. An analogous expression results for the
local three-body part of the correlated kinetic energy
$\corr{\uO}^{[3]}$, which involves \eqref{eq:ucom_mb_corr_kinpot}
and \eqref{eq:ucom_corr_kinpot}.

\section{Optimal Correlation Function}
\label{sec:opt}

The only input needed to evaluate correlated quantities explicitly is
the correlation function $\Rpm(r)$. Since $\Rpm(r)$ reflects the
properties of the correlations induced by the two-body interaction it
can be extracted from the interacting two-body system.

\subsection{Mapping of the $E=0$ Scattering Solution}
\label{sec:liq_optcorr2}

\begin{figure}
  \includegraphics[width=\columnwidth]{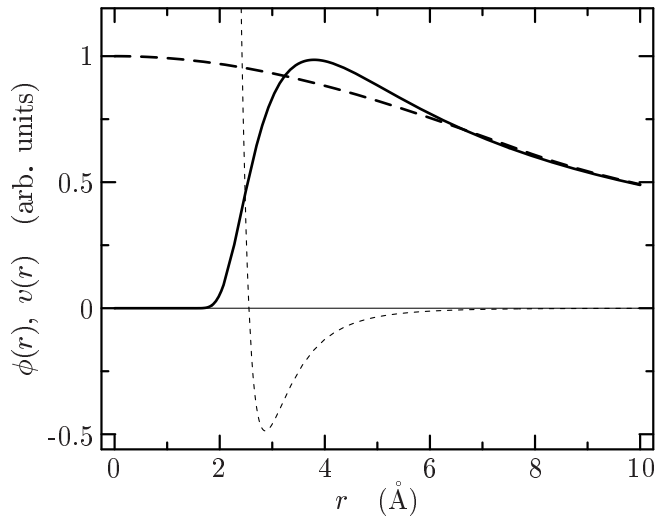}
  \caption{Exact relative wave function  $\phi_{\ex}(r)$ for the $E=0$
    scattering problem (solid line) in comparison with the hybrid
    ansatz $\phi_0(r)$ for $\xi=7.83\,\text{\AA}$ (dashed line). In 
    addition the radial dependence of the Lennard-Jones potential 
    $v(r)$ is shown (thin dotted line). }
  \label{fig:opt_E0solution}
\end{figure}

One method to construct an optimal correlation function is based on
the exact solution of the two-body Schr\"odinger equation. The exact
relative wave function $\phi_{\ex}(r)$ contains all information on
interaction induced two-body correlations. For the case of energy
$E=0$ and relative angular momentum $l=0$ the solid line in Figure
\ref{fig:opt_E0solution} shows the radial wave function (solid line)
for the Lennard-Jones potential \eqref{eq:intro_lennardjones} (thin
dotted line). The short-range part is dominated by a huge correlation
hole, i.e., a region where the strong repulsive core of the potential
enforces a vanishing wave function.

We can extract a correlation function that describes these short-range
correlations by requiring that the exact solution is reproduced by a
correlated ansatz wave function
\eq{
  \matrixe{\rV}{\CO}{\phi_0} \overset{!}{=} \braket{\rV}{\phi_{\ex}}
  \quad\text{or}\quad 
  \braket{\rV}{\phi_0} \overset{!}{=} \matrixe{\rV}{\adj{\CO}}{\phi_{\ex}} 
}
for all $r$ inside some maximum radius $\lambda$. Thus the optimal
correlation operator maps the short-range part of the uncorrelated
wave function $\phi_0(r)\equiv\braket{\rV}{\phi_0}$ onto the exact
solution $\phi_{\ex}(r)\equiv\braket{\rV}{\phi_{\ex}}$.  In coordinate
representation this condition leads with
\eqref{eq:ucom_corr_wavefct_2} and \eqref{eq:ucom_corr_metric} to
\eq{
  r\, \phi_0(r) 
  = \Rp(r) \sqrt{\Rp'(r)}\,\phi_{\ex}(r) ,
}
which can be written as an implicit integral equation for the optimal
correlation function $\Rp(r)$ \footnote{In the case of the strong core of
the Lennard-Jones potential it is numerically easier to solve the
integral equation for the inverse correlation function $\Rm(r)$.}
\eq{ \label{eq:opt_integraleq}
  [\Rp(r)]^3 
  = 3 \int_0^r\!\! \dd\xi\;\xi^2
    \bigg[ \frac{\phi_{0}(\xi)}{\phi_{\ex}[\Rp(\xi)]} \bigg]^2 .
}
The uncorrelated wave function $\phi_0(r)$ should be chosen in
accordance with the uncorrelated many-body trial state. For example,
if a condensate of $N$ uncorrelated bosons is described by a $N$-fold
product of a Gaussian-shaped single-particle wave functions the
relative wave function for each pair of bosons is again of Gaussian shape.

The choice of the uncorrelated wave function $\phi_0(r)$ determines,
which structures of the exact solution are generated by the
correlation operator and which have to be described by the ansatz
state. In order to isolate short-range correlations the uncorrelated
state should be able to describe the long-range behavior of the exact
solution.

According to this requirement we construct a hybrid wave function with
a long-range behavior given by the exact zero-energy solution outside
the range $\Lambda$ of the interaction:
\eq{ \label{eq:opt_hybrid_longrange}
  \phi_{\ex}(r) = 1 - a/r \quad\text{for}\;r>\Lambda,
}
where $a$ is the s-wave scattering length of the interaction. For two
${}^4$He atoms interacting via the Lennard-Jones potential
\eqref{eq:intro_lennardjones} we get $a = -171.3\,\text{\AA}$. The
short-range part of the ansatz state is described by a Gaussian. At
some radius $\chi$ both functions are matched with continuous
derivative. The final form of the uncorrelated wave function reads
\eq{ \label{eq:opt_hybrid}
  \phi_0(r) 
  = \begin{cases} 
  \frac{\chi-a}{\chi}
    \exp\!\big[ \frac{a}{2 (\chi-a)} \big(\tfrac{r^2}{\chi^2} - 1\big) \big]
    & ; r \le \chi \\[3mm]
  1 - a/r
    & ; r > \chi 
  \end{cases} . 
}
The matching radius $\chi$ is chosen such that the correlation
function $\Rp(r)-r$ determined from \eqref{eq:opt_integraleq} vanishes
at some finite radius $\lambda$. This allows a natural separation of
long- and short-range correlations in the two-body system.

\begin{figure}
  \includegraphics[width=\columnwidth]{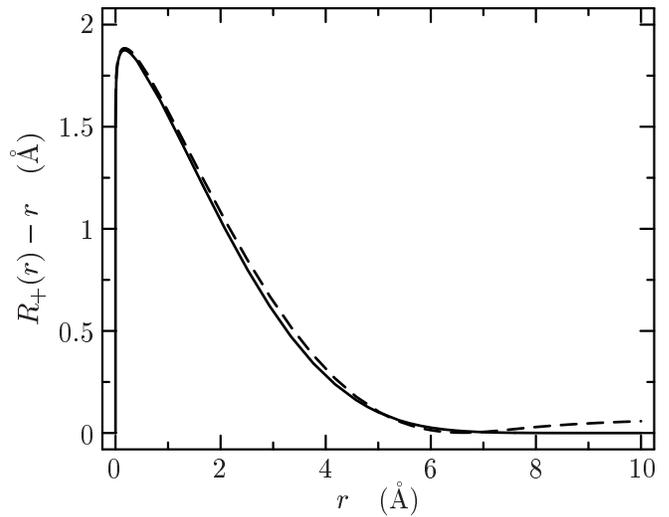}
  \caption{Optimal Correlation Function $\Rp(r)-r$ resulting from the
    integral equation \eqref{eq:opt_integraleq} (solid line) and the
    fitted parameterization (thin solid line). For comparison the
    parameterization with parameters resulting from energy
    minimization (thin dotted line) is depicted. }
  \label{fig:opt_correlator}
\end{figure}

In Figure \ref{fig:opt_E0solution} the hybrid ansatz with matching
radius $\chi=7.83\,\text{\AA}$ is shown in comparison with the exact
$E=0$ solution. Both wave functions agree nicely at large
distances. At small distances the exact solution shows the pronounced
depletion within the core and an enhancement in the attractive region
of the potential. These short-range correlations are not contained in
the uncorrelated wave function (dashed curve). Instead the unitary
correlation operator $\CO$, which maps the uncorrelated wave function
onto the exact eigenfunction (solid curve), takes over the task. As
discussed in section \ref{sec:ucom_operator} this has the great
advantage to allow the definition of correlated interactions that act
between simple uncorrelated many-body states like product states.

The solution of the integral equation \eqref{eq:opt_integraleq} for
the correlation function $\Rp(r)$ using these two wave functions is
shown by the dashed line in Figure \ref{fig:opt_correlator}. The
quantity $\Rp(r)-r$ corresponds to the net distance by which two
particles with an initial separation $r$ are shifted away from each
other. The shape of the curve shows that particles with distances
smaller than the core radius ($r_{\text{core}}\approx 2.5$\AA) are
shifted out of the core. If the initial distance is already larger
than the core radius, then the shift decreases rapidly and finally
vanishes at $r\approx6.6$\AA.  The small shift for larger radii
($r>6.6$\AA) originates from the minimal deviations of exact wave
function and hybrid ansatz in this range.  However these are long
range correlations, which should not be described by the unitary
correlation operator. Therefore we will set the function $\Rp(r)-r$ to
zero for $r\ge6.6$\AA\ in order to define the short-range correlation
function.

This intrinsic separation between long- and short-range correlations
works only if the potential has a sufficiently strong attractive
region in addition to the repulsive core. For this class of
interactions the exact two-body wave function is depleted inside the
core and enhanced in the attractive region. Since the unitary
correlation operator conserves the norm of the wave function it can
generate this kind of structure very easily: The amplitude shifted out
of the core is placed in the attractive region.

For purely repulsive potentials the situation is different. The exact
wave function of a scattering state is depleted inside the core and
the excess probability is spread over a large volume. Therefore the
unitary mapping leads to a correlator of large range. In order to stay
within the two-body approximation one has to restrict the range of the
correlations and compensate by improving the uncorrelated trial
states.

\subsection{Energy Minimization}
\label{sec:opt_corr1}

An alternative method to determine the correlation function $\Rp(r)$
emerges from the variational principle.  If we think of the energy
expectation value to be calculated with correlated states and the
uncorrelated Hamiltonian then the correlation function introduces
additional degrees of freedom into the trial state. By choosing a
suitable parameterization for $\Rp(r)$ with few variational parameters
we can minimize the correlated energy expectation value and obtain an
optimal correlation function.

According to the structure of the correlation function determined by
the mapping procedure described in the previous section we use the
following parameterization
\eq{ \label{eq:opt_corrpara}
  \Rp(r) = r + \alpha \; (r/\beta)^{\eta}\; \exp[-\exp(r/\beta)]
}
with three free parameters. The parameter $\beta$ determines the range
of the correlation function, $\alpha$ is related the maximum shift
distance and thus the radius of the core, and $\eta$ influences the
slope of $\Rp(r)$ at small radii, i.e., the ``hardness'' of the core.
We used this parameterization successfully for the description of the
structure of atomic nuclei and nuclear matter \cite{UCOM98,Roth00}.

In order to describe a homogeneous ${}^4$He liquid we choose a
constant wave function as two-body trial state $\braket{\rV}{\phi_0} =
1$. In that case only the local potentials $\corr{v}(\rO)$ and
$\corr{u}(\rO)$ contribute to the expectation value of the correlated
Hamiltonian
\eqmulti{
  E = \matrixe{\phi_0}{\corr{\HO}}{\phi_0} 
    &= \matrixe{\phi_0}{\corr{v}(\rO)+\corr{u}(\rO)}{\phi_0} \\ 
    &= \int\!\!\dd^3r\; [ \corr{v}(r) + \corr{u}(r) ] .
}
The expectation values of all momentum-dependent contributions
$t_{\rel},\corr{t}_{\Omega},\corr{t}_{r}$ in \eqref{eq:ucom_corr_kin}
vanish.  By inserting the parameterization of the correlation function
into \eqref{eq:ucom_corr_pot} and \eqref{eq:ucom_corr_kinpot} we can
express the integrand as function of the three variational parameters
$\alpha, \beta$ and $\eta$. The integration as well as the
minimization can easily be done numerically. We obtain a unique set of
parameters for the optimal correlation function
\eq{ \label{eq:opt_para}
\alpha = 6.267\,\text{\AA},\quad
\beta  = 3.520\,\text{\AA},\quad
\eta   = 0.052 .
}

The solid line in Figure \ref{fig:opt_correlator} shows the resulting
correlation function. It is in very good agreement with the
correlation function determined by the mapping procedure (dashed
line). This demonstrates that the parameterization used in the
variation is well suited for the correlation function.

For numerical simplicity we will use the parameterized correlation
function \eqref{eq:opt_corrpara} with parameters according to
\eqref{eq:opt_para} for the following investigations.

\subsection{Correlated and Effective Interaction}
\label{sec:opt_corrint}

\begin{figure}
  \includegraphics[width=\columnwidth]{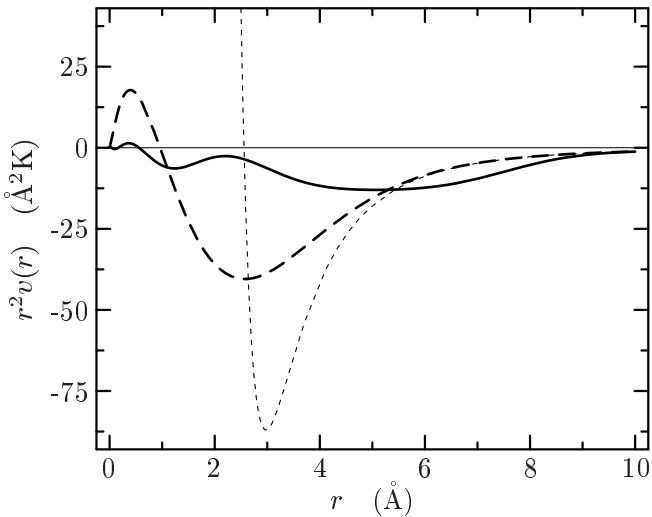}
  \caption{Local components of the correlated Hamiltonian: 
    uncorrelated Lennard-Jones potential $r^2 v(r)$ (thin dotted
    line), correlated potential $r^2 \corr{v}(r)$ (dashed line) and
    sum of correlated potential and local part of the correlated 
    kinetic energy $r^2 (\corr{v}(r)+\corr{u}(r))$ (solid line). }
  \label{fig:opt_corrint_local}
\end{figure}
\begin{figure}
  \includegraphics[width=\columnwidth]{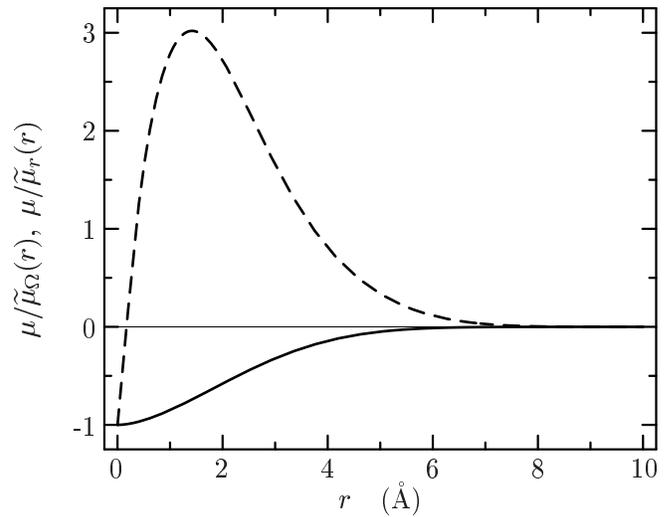}
  \caption{Radial dependencies of the effective mass corrections:
    angular effective mass correction $\mu/\corr{\mu}_{\Omega}(r)$
    (solid line) and radial effective mass correction 
    $\mu/\corr{\mu}_r(r)$ (dashed line). }
  \label{fig:opt_corrint_effmass}
\end{figure}

With the explicit form of the correlation function we can
study the structure of the correlated Hamiltonian in detail.
As shown in section \ref{sec:ucom_operator} the unitary transformation
generates a momentum-dependent effective interaction. 

The local part of this interaction consists of the correlated
interaction potential $\corr{v}(r)$ \eqref{eq:ucom_corr_pot} and a
local contribution $\corr{u}(r)$ \eqref{eq:ucom_corr_kinpot} from the
correlated kinetic energy operator. Figure \ref{fig:opt_corrint_local}
compares the radial dependencies of the uncorrelated Lennard-Jones
potential, the correlated Lennard-Jones potential and the sum of all
local contributions.  The unitary transformation of the potential
shifts the repulsive core to smaller radii. This leads to a
substantial reduction of the strength of the core, only a moderate
repulsion at very short ranges remains. If we add the local part of
the correlated kinetic energy the repulsive part vanishes completely;
the local component of the correlated Hamiltonian is purely
attractive.

Actually, it can be shown analytically that mapping of an uncorrelated
wave function, which is constant within the range of the repulsive
core, onto the exact $E=0$ solution results in a local part
$\corr{v}(r)+\corr{u}(r)$ of the correlated Hamiltonian that vanishes
identically inside the core \cite{UCOM98}. The small contributions of
the correlated local potentials within the core range that show up in
Figure \ref{fig:opt_corrint_local} result from the use of the
restricted parameterization \eqref{eq:opt_corrpara} for the
correlation function and the slight variation of the trial state.

The momentum-dependent parts of the correlated Hamiltonian, which are
formulated in terms of distance-dependent effective mass corrections
\eqref{eq:ucom_corr_effmass}, are shown in Figure
\ref{fig:opt_corrint_effmass}. The radial effective mass correction
(dashed line) generates a repulsion if the uncorrelated relative wave
function has non-vanishing radial derivatives. The angular effective
mass correction (solid line) generates an attraction if the uncorrelated
wave function has non-vanishing relative angular momentum. In general
the contribution of these components is small compared to the total
energy expectation value as the uncorrelated states have small
gradients.

Formally we can combine all the two-body terms of the correlated
Hamiltonian to a momentum-dependent effective interaction
\eq{
  \corr{\wO}^{[2]} 
  = \corr{v}(\rO) + \corr{u}(\rO) 
  + \corr{\tO}_{\Omega} + \corr{\tO}_r 
}
with a purely attractive local part given by \eqref{eq:ucom_corr_pot}
and \eqref{eq:ucom_corr_kinpot} and a momentum dependence given
by \eqref{eq:ucom_corr_effmasscorr}. With this effective interaction
the correlated Hamiltonian in two-body approximation has the
standard form
\eq{
  \corr{\HO}^{C2} = \TO + \sum_{i<j}^N \corr{\wO}^{[2]}_{ij} .
}
The next orders of the cluster expansion contribute additional
three-body, four-body, and higher order effective interactions.

The tamed effective two-body interaction $\corr{\wO}^{[2]}$ has a
different operator structure and completely different radial
dependencies than the uncorrelated potential $v(\rO)$. However, both
interactions are phase-shift equivalent, i.e., in a two-body
scattering process both interactions generate identical phase-shifts
for any collision energy, by construction. This generic property of
the effective interaction results from the finite range of the unitary
transformation, which does not influence the asymptotic behavior of
the scattering solutions. It is completely independent of the shape of
the correlation function at short ranges. In that way the unitary
correlator provides a recipe to generate an infinite manifold of
phase-shift equivalent interactions.  This aspect is of special
interest in cases where the interaction is determined only on the
basis of two-body scattering data --- like the nucleon-nucleon
interaction.

\section{Homogeneous ${}^4$He-Liquid}
\label{sec:liq}

As a first application of the UCOM formalism we want to investigate
the equation of state, i.e., the energy per particle as function of
the density, for a homogeneous ${}^4$He liquid at temperature $T=0$K.

In this case all effects that the interaction has on the many-body
state are described by the unitary correlation operator. The
uncorrelated many-body state $\ket{\Psi_0}$ is a direct product of
identical constant one-body states $\ket{\psi_0}$
\eq{ \label{eq:liq_mb_trialstate}
  \ket{\Psi_0} 
  = \ket{\psi_0}\otimes \cdots \otimes \ket{\psi_0} .
}
For the calculation of expectation values we assume a finite volume
$V$ containing $N$ particles. Accordingly the constant one-body states
read
\eq{
  \braket{\xV}{\psi_0} = V^{-1/2} .
}
The limit $N,V\to\infty$ at constant density $\rho=N/V$ is performed
in the final step of the calculation.

\subsection{Equation of State in Two- and Three-Body Approximation}
\label{sec:liq_eos}

First we calculate the energy per particle with the correlated
Hamiltonian in two-body approximation
\eq{ \label{eq:liq_hamiltonian_2b}
  \corr{\HO}^{C2} 
   = \TO + \corr{\VO}^{[2]} + \corr{\UO}^{[2]} + \corr{\TO}_{\Omega}^{[2]} + 
     \corr{\TO}_r^{[2]} .
}
The energy per particle in two-body approximation is defined by the
expectation value of this Hamiltonian calculated with the uncorrelated
many-body state \eqref{eq:liq_mb_trialstate}
\eq{
  \corr{\varepsilon}^{C2}(\rho) 
  = \frac{1}{N}\matrixe{\Psi_0}{\corr{\HO}^{C2}}{\Psi_0} .
}
Since the uncorrelated state is a product of constant one-body states
only the local components $\corr{\VO}$ and $\corr{\UO}$ contribute;
the expectation values of all momentum-dependent terms vanish.  A
simple calculation for the expectation value of the local components
leads to the following expression for the energy per particle in
two-body approximation
\eq{ \label{eq:liq_energy_2b}
  \corr{\varepsilon}^{C2}(\rho) 
  = (C_v^{[2]} + C_u^{[2]})\;\rho 
}
with constant coefficients given by
\eq{
  C_v^{[2]} = \frac{1}{2} \int\!\!\dd^3r\; \corr{v}(r) ,
  \qquad
  C_u^{[2]} = \frac{1}{2} \int\!\!\dd^3r\; \corr{u}(r) .
}
The integrals over the correlated potential $\corr{v}(r)$ given by
\eqref{eq:ucom_corr_pot} and the local part of the correlated kinetic
energy $\corr{u}(r)$ according to \eqref{eq:ucom_corr_kinpot} are
evaluated numerically
\eq{ \label{eq:liq_energy_2b_coeff}
  C_v^{[2]} = -856.5 \, \text{K\AA}^3, 
  \qquad
  C_u^{[2]} = 401.5 \, \text{K\AA}^3 .
}

Obviously, for homogeneous Bose fluids at zero temperature the
two-body approximation is not able to describe saturation, i.e., a
minimum of the energy at some finite density. The energy in two-body
approximation \eqref{eq:liq_energy_2b} is always proportional to
density and due to the negative sum of the coefficients
\eqref{eq:liq_energy_2b_coeff} drops with increasing density.

\begin{figure}
  \includegraphics[width=\columnwidth]{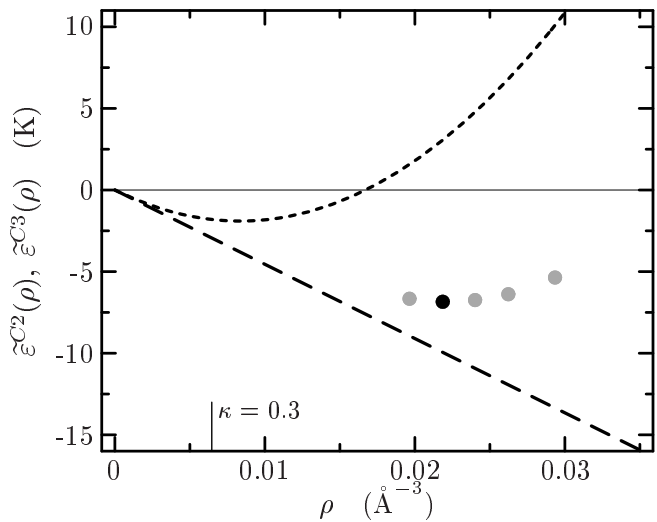}
  \caption{Energy per particle for a homogeneous ${}^4$He liquid as
    function of density calculated in two-body approximation
    $\corr{\varepsilon}^{C2}(\rho)$ (dashed line) and in three-body
    approximation $\corr{\varepsilon}^{C3}(\rho)$ (dotted line).  The
    vertical line marks the density at which the smallness parameter
    $\kappa$ reaches the value $0.3$. The circles show the results of
    an exact GFMC calculation \cite{KaLe81}; statistical errors are
    smaller than the size of the symbols. }
  \label{fig:liq_energy_2b3b}
\end{figure}

The dashed line in Figure \ref{fig:liq_energy_2b3b} shows this
behavior in comparison with results of a Green's Function Monte Carlo
(GFMC) calculation \cite{KaLe81}, which are exact up to the
statistical errors of the Monte Carlo sampling. Despite the fact that
the two-body approximation fails to generate a minimum, the predicted
orders of magnitude for the energy are correct. This is remarkable,
considering that the uncorrelated Lennard-Jones potential would give
an infinite positive energy expectation value for all densities. The
special unitary transformation that we apply tames the strong core of
the potential and leads to a bound many-body system.

We note that the two-body approximation does generate saturation in
other systems like homogeneous Fermi liquids. In these cases the mass
corrections contribute, because of the Fermi motion, and generate a
positive contribution to the energy that grows with a higher power in
density than the attractive local terms.

A straightforward step to go beyond the two-body approximation is the
inclusion of the three-body terms of the cluster expansion. In
addition to the terms in \eqref{eq:liq_hamiltonian_2b} the correlated
Hamiltonian in three-body approximation contains three-body
contributions from the correlated potential and the correlated kinetic
energy
\eq{ \label{eq:liq_hamiltonian_3b}
  \corr{\HO}^{C3} 
   = \corr{\HO}^{C2} + \corr{\VO}^{[3]} + \corr{\UO}^{[3]} 
   + \corr{\TO}^{[3]} .
}
Again only the local three-body terms $\corr{\VO}^{[3]}$ and
$\corr{\UO}^{[3]}$ contribute to the energy expectation value with the
uncorrelated states \eqref{eq:liq_mb_trialstate}. The resulting energy
per particle in three-body approximation reads
\eq{ \label{eq:liq_energy_3b}
  \corr{\varepsilon}^{C3}(\rho) 
  = (C_v^{[2]} + C_u^{[2]})\;\rho + (C_v^{[3]} + C_u^{[3]})\;\rho^2 .
}
The coefficients are defined by the six-dimensional integrals 
\eq{
  C_{v}^{[3]} 
  = \frac{1}{6}\int\!\!\dd^3x_1\dd^3x_2\;\corr{v}^{[3]}(\xV_1,\xV_2,\xV_3=0) 
}
and the corresponding expression for $C_{u}^{[3]}$. The integration is
done with a standard Monte Carlo algorithm (``VEGAS'' from
\cite{PrTe92}) and yields
\eqmulti{
  C_v^{[3]} &= (-2512 \pm 50)\, \text{K\AA}^6, \\
  C_u^{[3]} &= (29739 \pm 1100)\, \text{K\AA}^6 .
}
The local three-body terms of the correlated Hamiltonian generate a
positive contribution to the energy per particle which is proportional
to $\rho^2$. The dominant term is the local three-body contribution of
the correlated kinetic energy. Together with the negative
density-proportional contribution of the local two-body potentials the
energy in three-body approximation \eqref{eq:liq_energy_3b} provides a
minimum at finite density.

The density dependence of the energy per particle in three-body
approximation $\corr{\varepsilon}^{C3}(\rho)$ is shown by the dotted
curve in Figure \ref{fig:liq_energy_2b3b}. Obviously the three-body
contribution is not a small correction to the two-body approximation
for densities larger than $0.005\,\text{\AA}^{-3}$, which is only one
fourth of the saturation density. This demonstrates that many-body
correlations have a strong influence in these systems and that the
two-body approximation alone is not sufficient to describe ground
state properties. A similar conclusion can be drawn from the smallness
parameter $\kappa = \rho\,V_C$. It reaches the phenomenological limit
$\kappa=0.3$ for the validity of the two-body approximation at about
$1/3$ of the expected saturation density .

However, the naive inclusion of the three-body terms of the cluster
expansion of the correlated Hamiltonian does not improve the result
beyond a density of about $0.005\,\text{\AA}^{-3}$. Although we obtain
saturation in three-body approximation, both, energy and density at
the minimum of $\corr{\varepsilon}^{C3}(\rho)$ are substantially
smaller than the results of the GFMC calculation. This discrepancy has
several reasons: Firstly, the optimal correlation function was
determined in the two-body system, i.e., in two-body
approximation. For a consistent treatment on the level of the
three-body approximation the optimal correlation function should be
determined by minimizing the energy in three-body
approximation. Secondly, we expect that genuine three-body
correlations play a very strong role \cite{UsFa82,Pand78}. They are
described by an irreducible three-body part $\sum_{i<j<k} \gO_{ijk}$
in the generator \eqref{eq:ucom_generator} that was not included in
the present treatment. Finally, in view of the large three-body
contributions there is no good reason to neglect terms beyond
three-body order.

Each of the points mentioned can in principle be explicitly included
in the calculation. However, none of these corrections are expected to
yield reasonably converged results in the particular case of the
${}^4$He liquid. The density range where the description should be
applied is far above the range where the two-body approximation is
applicable, i.e., where the smallness parameter $\kappa$ is below the
typical value of $0.3$. The successive inclusion of higher orders of
the cluster expansion will extend this range but one would have to
include very high orders --- which are beyond the numerical
possibilities --- to reach the expected saturation density. The only
reasonable way to extend the treatment is a partial summation over all
cluster orders. This is done in the framework of Jastrow correlations
with the so called Hypernetted Chain (HNC) summation schemes
\cite{PaSc77,Pand78,Clar79,UsFa82}. However, this type of partial
summation is not feasible for the unitary correlation operator because
of its more complicated operator structure.

\subsection{Equation of State with Density-Dependent 
  Correlation Functions}
\label{sec:liq_eos_densdep}

We aim at an effective description of the higher orders of the cluster
expansion that still allows a compact analytic formulation of
correlated observables and does not require extensive numerical efforts. 
One general way to simulate the effect of higher cluster orders is
to introduce density-dependent correlation functions on the level of
the two-body approximation.

At low densities the correlator shifts a pair of particles in an
optimal way from the repulsive into the attractive region. When the
density grows other particles will be nearby and obstruct the
correlation that was optimal for a free particle pair. To simulate
this effect the range of the correlation function $\Rp(r)$ should be
more and more reduced for increasing density. This will reduce the
binding as the particles will not be able any more to fully exploit
the attractive region of the potential.

In practice this is accomplished by scaling the parameters $\alpha$
and $\beta$, which have the dimension of a length, in the
parameterization \eqref{eq:opt_corrpara} of the correlation function
$\Rp(r)$ with some factor $\xi(\rho) \lesssim1$ that depends on
density. At very low densities we assume that the higher-order
contributions are negligible such that $\xi(\rho\to0)=1$. With growing
density the effect of the higher cluster orders grows such that the
scaling factor $\xi(\rho)$ has to drop. The most simple ansatz for the
density dependence of $\xi(\rho)$ is
\eq{ \label{eq:opt_scale_densdep}
  \xi(\rho) = 1 - \gamma \rho,
}
with one free parameter $\gamma$. The density-dependent correlation
function reads
\eq{ \label{eq:opt_corrpara_densdep}
  \Rp(r,\rho) 
  = r + \alpha\xi(\rho)  \Big[\frac{r}{\beta\xi(\rho)}\Big]^{\!\eta} 
    \exp\!\!\Big[\!\!-\exp\!\Big(\frac{r}{\beta\xi(\rho)}\Big)\!\Big].
}
Using this correlation function all components of the correlated
Hamiltonian --- like correlated potential, local part of the
correlated kinetic energy, and effective mass corrections --- become
explicitly density-dependent. The energy per particle is given by
\eq{ \label{eq:liq_energy_densdep}
  \corr{\varepsilon}^{C2}_{\rho}(\rho) 
  = \frac{\rho}{2} \int\!\!\dd^3r\;
    [\corr{v}_{\rho}(r,\rho) + \corr{u}_{\rho}(r,\rho)],
}
where the subscript $\rho$ indicates that the density-dependent
correlation function is used. 

Due to the explicit density-dependence of the correlated local
potentials the equation of state is not proportional to $\rho$ any
more. From a Taylor expansion of \eqref{eq:liq_energy_densdep} around
$\rho=0$ we obtain terms to all powers of the density. Thus
all orders of the cluster expansion are represented by the
density-dependent correlator in an effective way.  

The phenomenological parameter $\gamma$ in the scaling factor
\eqref{eq:opt_scale_densdep} is in general adjusted such that
\eqref{eq:liq_energy_densdep} agrees with one experimental data point,
or for some density with the result of a realistic calculation. We
will use the result of the GFMC calculation \cite{KaLe81} for the
density $\rho^{\text{GFMC}}_{\sat}$ and energy
$\varepsilon^{\text{GFMC}}_{\sat}$ at the saturation point and require
\eq{
  \corr{\varepsilon}^{C2}_{\rho}(\rho^{\text{GFMC}}_{\sat}) 
  \overset{!}{=} \varepsilon^{\text{GFMC}}_{\sat} .
}
This ensures that the equation of state \eqref{eq:liq_energy_densdep}
calculated with the density-dependent correlator runs through that
point. However, this does not imply that the minimum of
\eqref{eq:liq_energy_densdep} coincides with the GFMC minimum. Using
the values \cite{KaLe81}
$\rho^{\text{GFMC}}_{\sat}=0.0219\,\text{\AA}^{-3}$ and
$\varepsilon^{\text{GFMC}}_{\sat}=-6.848\,\text{K}$ we obtain
\eq{ \label{eq:opt_densdep_para}
  \gamma = 3.696\,\text{\AA}^3 .
}
We have chosen to fix the density dependence, i.e., $\gamma$, with a
calculation instead of an experimental value, because that calculation
uses the same potential \eqref{eq:intro_lennardjones} and it is
known that this simple Lennard-Jones potential does not reproduce
exactly the experimental two-body data.

\begin{figure}
  \includegraphics[width=\columnwidth]{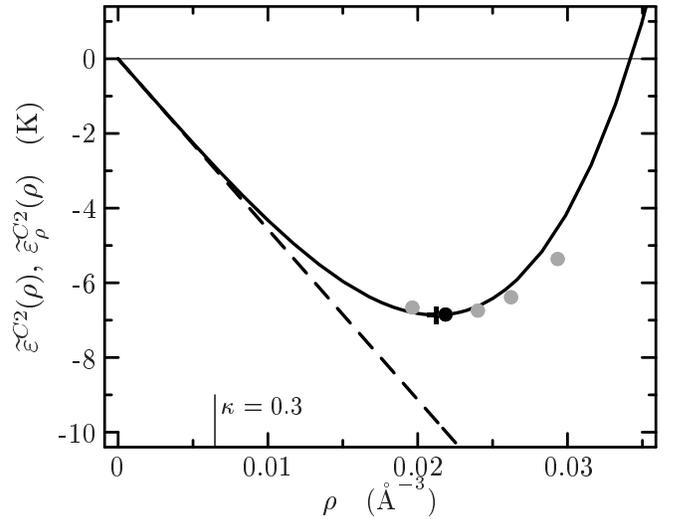}
  \caption{Energy per particle for a homogeneous ${}^4$He liquid
    calculated in two-body approximation with a two-body optimized
    correlation function $\corr{\varepsilon}^{C2}(\rho)$ (dashed line)
    and with density-dependent correlation function
    $\corr{\varepsilon}^{C2}_{\rho}(\rho)$ (solid line). The vertical
    line marks the density at which the smallness parameter $\kappa$
    reaches the value $0.3$.  The gray circles show the results of an
    exact GFMC calculation \cite{KaLe81}, the minimum is indicated by
    the black circle.  }
  \label{fig:liq_energy_2bdd}
\end{figure}

Figure \ref{fig:liq_energy_2bdd} shows the energy per particle
$\corr{\varepsilon}^{C2}_{\rho}(\rho)$ calculated with the
density-dependent correlation function.  The phenomenological
description of the higher order contributions by a density-dependent
correlation function indeed generates a minimum of the energy with
\eq{
  \rho_{\sat} = 0.0212\,\text{\AA}^{-3},
  \qquad
  \corr{\varepsilon}^{C2}_{\rho,\sat}=-6.860\,\text{K} .
}
The position of the minimum (indicated by the cross) as well as the
shape of the equation of state around the minimum agree very nicely
the the GFMC result. As noted above this is not an immediate
consequence of the adjustment of $\gamma$. The agreement shows that
the phenomenological density-dependence is well suited to describe the
major aspects of many-body correlations on a numerically very simple
level. Actually, all calculations have been performed within a simple
\emph{Mathematica} notebook.

Once the density-dependence is fixed the correlation function can be
used to evaluate any correlated observable of interest. The analytic
expressions for many correlated observables in two-body approximation
are rather simple, but the evaluation of higher cluster orders becomes
complicated if not intractable. Thus the inclusion of the
phenomenological treatment of higher order effects by means of the
density-dependent correlator is a very useful tool.  Interesting
quantities are, e.g., the radial two-body distribution function
$g(r)$, the static structure factor $S(q)$ or the one-body momentum
distribution $n(k)$. The corresponding observables for fermionic
liquids were discussed elsewhere \cite{Roth00}.

In the following section we will use the density-dependent correlation
function to investigate the structure of small droplets of
${}^4$He. This will demonstrate the universality of the concept.

\section{Small ${}^4$He-Droplets}
\label{sec:drop}

In a next step we want to use the Unitary Correlation Operator Method
to investigate the ground state properties of small ${}^4$He droplets.

The many-body problem is treated in a simple variational framework. The
many-body trial state $\ket{\Psi,N}$ is --- like for the homogeneous
liquid --- assumed to be a direct product of identical one-body
states. The one-body states are described by a Gaussian wave function
\eq{
  \braket{\vec{x}}{\psi} 
  = (\pi a)^{-3/4} \exp[-\xV^2/(2 a)] 
}
with a common width parameter $a$ as the only variational degree of
freedom.  The Gaussian trial state is applicable for small droplets
with $N\lesssim70$; larger droplets show a saturation of the central
density, which cannot be modeled with this ansatz \cite{PaPi86,KrCh95}.
An advantage of the Gaussian ansatz is that the two-body wave function
\eq{
  \braket{\xV_1,\xV_2}{\psi,\psi} 
  = \braket{\XV}{\Phi} \braket{\rV}{\phi}
}
can be separated analytically into a center of mass component
$\braket{\XV}{\Phi}$ and a relative wave function $\braket{\rV}{\phi}$
\eqmulti{ \label{eq:drop_2b_wavefunc_sep}
  \braket{\XV}{\Phi} &= (\pi a/2)^{-3/4} \exp[-\XV^2\!/a] ,\\
  \braket{\rV}{\phi} &= (2\pi a)^{-3/4} \exp[-\rV^{\,2}\!/(4 a)] .
}

This simple trial state is able to describe mean-field like effects of
the interaction, i.e., the spatial localization of the atoms. However,
any correlation effect beyond the mean-field level has to be described
by the unitary correlation operator.

\subsection{Binding Energy and rms-Radius in Two-Body Approximation}

First we use the two-body approximation with the correlation function
\eqref{eq:opt_corrpara} obtained from the two-body scattering solution
to investigate the ground state properties of small droplets. The
correlated intrinsic Hamiltonian $\corr{\HO}^{C2}_{\intr} =
\corr{\HO}^{C2} - \TO_{\cm}$ in two-body approximation has the
structure
\eq{ \label{eq:drop_hamiltonian_2b}
  \corr{\HO}^{C2}_{\intr} 
   = \TO_{\intr} + \corr{\VO}^{[2]} + \corr{\UO}^{[2]} 
     + \corr{\TO}_{\Omega}^{[2]} + \corr{\TO}_r^{[2]} ,
}
where capital letters indicate summation over all particles, like
$\corr{\VO}^{[2]} = \sum_{i<j}^N \corr{\vO}^{[2]}_{ij}$. The
uncorrelated intrinsic kinetic energy $\TO_{\intr}$, i.e., the total
kinetic energy $\TO$ reduced by the center of mass contribution
$\TO_{\cm}$, can be expressed by the relative two-body kinetic energy
$\tO_{\rel}$ 
\eq{ \label{eq:drop_kin_int}
 \TO_{\intr} = \TO - \TO_{\cm}
 = \frac{2}{N} \sum_{i<j}^N \tO_{\rel,ij} ,
}
where
\eq{
  \TO
  = \frac{1}{2m}\sum_i^N\pOV_i^2 
  \;,\qquad
  \TO_{\cm}
  = \frac{1}{2mN}\bigg[\sum_i^N \pOV_i\bigg]^2 .
}
Due to the simple structure of the many-body trial state the
expectation value of the correlated Hamiltonian
\eqref{eq:drop_hamiltonian_2b} in the $N$-body system can be expressed
by that of the two-body system
\eqmulti{
  \corr{E}^{C2} 
  &= \matrixe{\Psi,N}{\corr{\HO}^{C2}_{\intr}}{\Psi,N}  \\
  &= \frac{N(N-1)}{2}\matrixe{\Psi,2}{\corr{\HO}^{C2}_{\intr}}{\Psi,2} .
}
Moreover we can use that the intrinsic Hamiltonian
\eqref{eq:drop_hamiltonian_2b} acts on the relative part
$\braket{\rV}{\phi}$ of the two-body wave function only. This leads to
a simple form of the correlated energy expectation value of the
$N$-body system
\eqmulti{ \label{eq:drop_energy_2b}
  \corr{E}^{C2} 
  &= (N-1)\;\matrixe{\phi}{\tO_{\rel}}{\phi} \\
  &+ \frac{N(N-1)}{2} \matrixe{\phi}{ \corr{v}(\rO) 
    + \corr{u}(\rO) + \corr{\tO}_r}{\phi} .
}
We used that the expectation value of the angular effective mass
correction $\corr{t}_{\Omega}$ vanishes due to the spherical symmetry
of the relative wave function (relative s-wave state). The two-body
expectation value of the relative kinetic energy is given by
\eq{
  \matrixe{\phi}{\tO_{\rel}}{\phi} 
  = \frac{1}{2\mu} \int\!\!\dd^3r\; \Big|
    \frac{\partial}{\partial r}\braket{r}{\phi} \Big|^2
  = \frac{1}{2\mu} \frac{3}{4a}.
}
For the correlated two-body potential and the local part of the
correlated kinetic energy we obtain
\eq{ \label{eq:drop_expect_pot}
  \matrixe{\phi}{\corr{v}(\rO) + \corr{u}(\rO)}{\phi} 
  = \int\!\!\dd^3r\; [\corr{v}(r)+\corr{u}(r)]\;\big|\braket{r}{\phi}\big|^2 .
}
Finally, the expectation value of the radial effective mass correction reads
\eq{ \label{eq:drop_expect_effmass}
  \matrixe{\phi}{\corr{\tO}_r}{\phi} 
  = \frac{1}{2\mu} \int\!\!\dd^3r\; \frac{\mu}{\corr{\mu}_r(r)} 
    \Big|\frac{\partial}{\partial r}\braket{r}{\phi} \Big|^2 .
}
For a fixed correlation function $\Rp(r)$ the radial dependencies of
the correlated potential $\corr{v}(r)$, the local part of the
correlated kinetic energy $\corr{u}(r)$, and the radial effective mass
correction $\mu/\corr{\mu}_r(r)$ are known analytically (see section
\ref{sec:ucom_operator}). The minimization of the energy
$\corr{E}^{C2}$ as function of the width parameter $a$ as well as the
calculation of the integrals in equations \eqref{eq:drop_expect_pot}
and \eqref{eq:drop_expect_effmass} is done numerically.

Another interesting observable is the mean-square radius of the
droplet, defined by the expectation value of the operator
\eq{
  \RO_{\text{ms}} 
  = \frac{1}{N} \sum_i^N (\xOV_i - \XOV_{\cm})^2 
  = \frac{1}{N^2} \sum_{i<j}^N \rO_{ij}^2 , 
}
where $\XOV_{\cm} = \frac{1}{N} \sum_i^N \xOV_i$ is the center of mass
coordinate of the many-body system. The formulation in terms of the
two-body operator $\rO^2$ reveals a similar structure like for the
intrinsic kinetic energy \eqref{eq:drop_kin_int}. The many-body
product state under consideration allows a direct calculation of the
uncorrelated expectation value using the relative two-body wave
function \eqref{eq:drop_2b_wavefunc_sep}
\eqmulti{ \label{eq:drop_msradius_expect}
  r_{\text{rms}}
  &= \big[ \matrixe{\Psi,N}{\RO_{\text{ms}}}{\Psi,N} \big]^{1/2} \\ 
  &= \bigg[\frac{N-1}{2N} \matrixe{\phi}{\rO^2}{\phi} \bigg]^{1/2}  
  = \bigg[\frac{N-1}{2N}\, 3 a\bigg]^{1/2} .
}
The correlated ms-radius $\corr{\RO}_{\text{ms}}^{C2}$ in two-body
approximation has a similar structure like the kinetic energy
\eq{ \label{eq:drop_corr_msradius}
  \corr{\RO}_{\text{ms}}^{C2} 
  = \RO_{\text{ms}} + \corr{\RO}_{\text{ms}}^{[2]} .
}
In addition to the uncorrelated mean-square radius $\RO_{\text{ms}}$
the unitary transformation generates a two-body contribution
\eq{
  \corr{\RO}_{\text{ms}}^{[2]}
  = \frac{1}{2N} \sum_{i<j}^N \big[\Rp^2(\rO_{ij}) - \rO_{ij}^2\big] .
}
The expectation value of the correlated ms-radius
\eqref{eq:drop_corr_msradius} can again be expressed by the relative
two-body wave function alone
\eq{ \label{eq:drop_corr_msradius_expect}
  \corr{r}_{\text{rms}}^{C2} 
  = \bigg[\frac{N-1}{2N} \matrixe{\phi}{\rO^2 
    + \frac{N}{2}\big(\Rp^2(\rO)-\rO^2\big)}{\phi} \bigg]^{1/2}.
}

In order to study the quality of the two-body approximation we first
look at very small droplets with $N\le10$. For these droplet sizes an
early variational Monte Carlo (VMC) calculation on the basis of the
Lennard-Jones potential exists \cite{ScSc65}. The authors use a
Jastrow-type parameterization of two-body correlations with a
long-range form adjusted to the behavior of the exact two-body
solution. The calculation of the many-body expectation value with
these trial states involves a Monte Carlo integration routine. The
resulting energy expectation values show a statistical error of
approximately 10\%.

\begin{figure}
  \includegraphics[width=0.9\columnwidth]{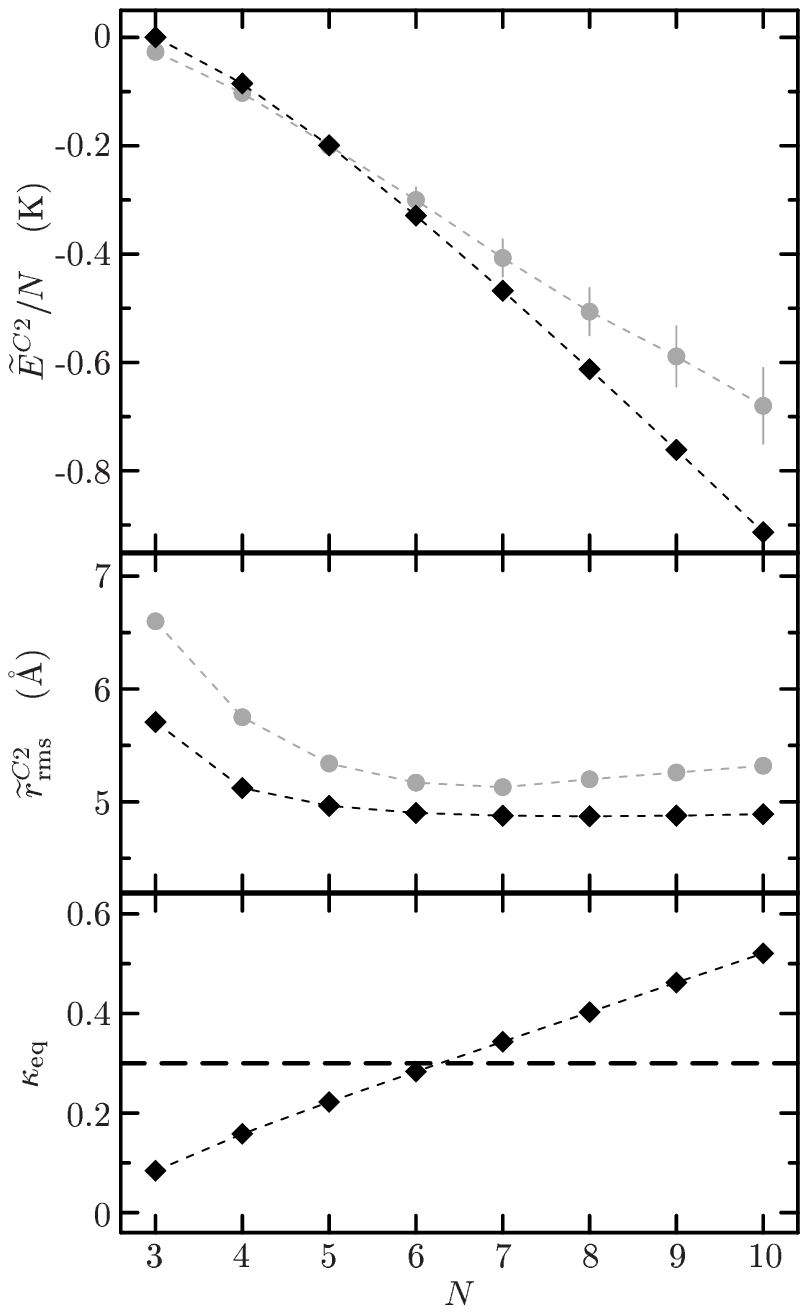}
  \caption{Energy per particle (upper panel), rms-Radius (middle panel),
    and smallness parameter $\kappa_{\equi}$ (lower panel) for small
    ${}^4$He droplets in two-body approximation with two-body optimized
    correlation function. The gray circles show results of a
    variational Monte Carlo calculation \cite{ScSc65}, statistical
    errors for the energy are indicated by the gray bars. The lines
    should guide the eye.  }
  \label{fig:drop_2b}
\end{figure}

Our results of the minimization of $\corr{E}^{C2}$
\eqref{eq:drop_energy_2b} with respect to $a$ using the system
independent correlation function given in equations
\eqref{eq:opt_corrpara} and \eqref{eq:opt_para} for $N\le10$ are
summarized in Figure \ref{fig:drop_2b}. The upper panel shows the
correlated energy per particle, the middle panel the correlated
rms-radius of the droplet and the lower panel the smallness parameter
$\kappa_{\equi}$ versus the number of particles $N$.

As for the homogeneous liquid the expectation values of the
uncorrelated Lennard-Jones potential would diverge. But the unitary
transformation of the Hamiltonian using the correlation function
determined in the two-body system removes the strong short-range
repulsion of the potential completely. With this tamed correlated
potential and a simple product ansatz for the many-body wave function
the energies of very loosely bound droplets actually agree very nicely
for $N\le6$ with the VMC result.  For $N>6$ the energy expectation
value is below the VMC value and the difference grows with increasing
particle number or density.  Since expectation values are evaluated in
two-body approximation, i.e., the full cluster expansion is truncated
above two-body order, a basic property of the Ritz variational
principle is lost: The variational minimum of the energy is not necessarily
bounded from below by the exact energy eigenvalue. The observed
overbinding indicates that the higher orders of the cluster expansion
have to give a sizeable positive contribution to raise the expectation
value above the exact eigenvalue.

In direct connection with the overbinding in two-body approximation
the correlated rms-radii of energy minimized states are systematically
smaller than the VMC result. Accordingly the densities are too high.
In contrast to the homogeneous liquid the two-body approximation does
not produce a pathological collapse towards high densities for small
droplets. This is due to the positive contributions of the kinetic
energy and the effective mass correction that are absent in the
homogeneous case. 
  
As a measure for the validity of the two-body approximation we used
the smallness parameter $\kappa = \rho V_{C}$ with correlation volume
$V_C$ defined by \eqref{eq:ucom_corrvol}. In order to specify the
smallness parameter for inhomogeneous systems we have to define a
measure for the average density. One possible definition is the value
of the density for a step-like density profile with same
(uncorrelated) rms-radius as the original distribution. For identical
Gaussian single-particle wave functions with width parameter $a$ this
equivalent density is given by
\eq{ \label{eq:drop_equivdensity}
  \rho_{\equi} 
  = \frac{3 N}{4 \pi} \bigg[ \frac{3}{5} \; 
    \frac{1}{r_{\text{rms}}^2} \bigg]^{3/2} 
  = \frac{3 N}{4 \pi} \bigg[\frac{2}{5} \; 
    \frac{N}{N-1}\; \frac{1}{a} \bigg]^{3/2}.
}
The smallness parameter $\kappa_{\equi} = \rho_{\equi} V_C$ defined
with this density is shown in the lower panel of Figure
\ref{fig:drop_2b} as function of particle number. Like for the
homogeneous liquid the energy in two-body approximation starts to
deviate from the exact result if the smallness parameter
$\kappa_{\equi}$ exceeds the value $0.3$ (indicated by the dashed
line). This again confirms that the two-body approximation is valid as
long as the smallness parameter fulfills the condition $\kappa
\lesssim 0.3$.

\subsection{Energy and rms-Radius with Density-Dependent Correlation Function}

To account for the effect of many-body correlations in a simple but
efficient way we employ the concept of density-dependent correlation
functions introduced in section \ref{sec:liq_eos_densdep} for the
homogeneous liquid.
 
To apply the density-dependent correlator in an inhomogeneous system
we have to specify in which way the density-dependence should be
evaluated. The most simple approach is to insert the equivalent
density of the droplet, as defined in \eqref{eq:drop_equivdensity},
into the density-dependent correlation function. That means we neglect
that the effects from stronger many-body correlations in the center of
the droplet do not exactly cancel the weaker ones at the surface.

A more elaborate ansatz to account for the inhomogeneity would be to
implement the density-dependence in local density approximation, i.e.,
the correlation function which acts on the relative coordinate of a
particle pair depends on the local density at their center of mass
position. This is not done here.
\setlength{\tabcolsep}{8pt}
\begin{table*} 
\caption{Energy expectation values and rms-radii for small ${}^4$He
  droplets with different particle number $N$ resulting from energy
  minimization in two-body approximation with two-body optimized and
  density-dependent correlation function. The width parameter $a$ is
  given in units of \AA${}^2$, energies are given per particle in units
  of K, and the rms-radius in units of \AA.} 
\label{tab:drop}
\begin{ruledtabular}
\begin{tabular}{c | c c c | c c c c c c c | c c}
  & \multicolumn{3}{c|}{Two-Body Approximation}
    & \multicolumn{7}{c|}{Two-Body Approximation with Density-Dependent Correlator}
      & \multicolumn{2}{c}{Ref. \cite{ScSc65}}
\\
$N$
  & $a$ & $\corr{E}^{C2}\!/N$ & $\corr{r}_{\text{rms}}^{C2}$
    & $a$ & $T_{\text{int}}/N$ & $\corr{T}_{r,\rho}^{[2]}/N$ &
    $\corr{U}^{[2]}_{\rho}/N$ & $\corr{V}^{[2]}_{\rho}/N$
    & $\corr{E}^{C2}_{\rho}\!/N$ & $\corr{r}_{\text{rms},\rho}^{C2}$
      & $E/N$ & $r_{\text{rms}}$

\\
\hline
3
  & 32.33 & 0.000 & 5.71   
    & 34.39 & 0.176 & 0.005 & 0.238 & -0.418 & 0.002 & 5.88
      & -0.027 & 6.60
\\
4
  & 22.83 & -0.085 & 5.12
    & 25.19 & 0.271 & 0.021 & 0.551 & -0.919 & -0.076 & 5.37
      & -0.103 & 5.75
\\
5
  & 19.81 & -0.199 & 4.97
    & 22.70 & 0.320 & 0.038 & 0.846 & -1.379 & -0.175 & 5.29
      & -0.200 & 5.34
\\
6
  & 18.26 & -0.329 & 4.90
    & 21.71 & 0.349 & 0.054 & 1.122 & -1.807 & -0.282 & 5.30
      & -0.300 & 5.17
\\
7
  & 17.31 & -0.468 & 4.88
    & 21.32 & 0.366 & 0.067 & 1.378 & -2.202 & -0.391 & 5.34
      & -0.407 & 5.13
\\
8
  & 16.67 & -0.613 & 4.87
    & 21.23 & 0.375 & 0.079 & 1.614 & -2.567 & -0.500 & 5.40
      & -0.506 & 5.20
\\
9
  & 16.20 & -0.761 & 4.88
    & 21.31 & 0.379 & 0.088 & 1.832 & -2.905 & -0.606 & 5.46
      & -0.589 & 5.26
\\
10
  & 15.84 & -0.913 & 4.89
    & 21.50 & 0.380 & 0.096 & 2.033 & -3.218 & -0.709 & 5.53
      & -0.680 & 5.32
\\
20
  & 14.42 & -2.500 & 5.14
    & 25.13 & 0.343 & 0.118 & 3.427 & -5.442 & -1.554 & 6.20
\\
30
  & 14.01 & -4.125 & 5.44
    & 29.15 & 0.301 & 0.109 & 4.221 & -6.778 & -2.146 & 6.75
\\
40
  & 13.81 & -5.760 & 5.74
    & 32.97 & 0.269 & 0.098 & 4.745 & -7.698 & -2.588 & 7.20
\\
50
  & 13.69 & -7.399 & 6.02
    & 36.59 & 0.243 & 0.085 & 5.119 & -8.381 & -2.933 & 7.60
\\
60
  & 13.61 & -9.040 & 6.29
    & 40.02 & 0.223 & 0.076 & 5.405 & -8.918 & -3.214 & 7.95
\\
70
  & 13.56 & -10.683 & 6.55
    & 43.29 & 0.207 & 0.068 & 5.631 & -9.353 & -3.447 & 8.26
\\
\end{tabular}
\end{ruledtabular}
\end{table*}

The variational calculation of the ground state properties of small
${}^4$He droplets including a correlation function that depends on the
equivalent density \eqref{eq:drop_equivdensity} is
straightforward. The parameterization \eqref{eq:opt_corrpara_densdep}
of the correlation function $\Rp(r,\rho_{\equi})$ is used with the
parameters \eqref{eq:opt_para} determined in the two-body system. The
parameter of the density-dependent scaling function
\eqref{eq:opt_scale_densdep} is taken from the investigations of the
homogeneous liquid \eqref{eq:opt_densdep_para}. The energy expectation
value in two-body approximation with density-dependent correlations is
of the same form as for the density-independent correlation functions
discussed in the preceding section.
\begin{figure}
  \includegraphics[width=0.9\columnwidth]{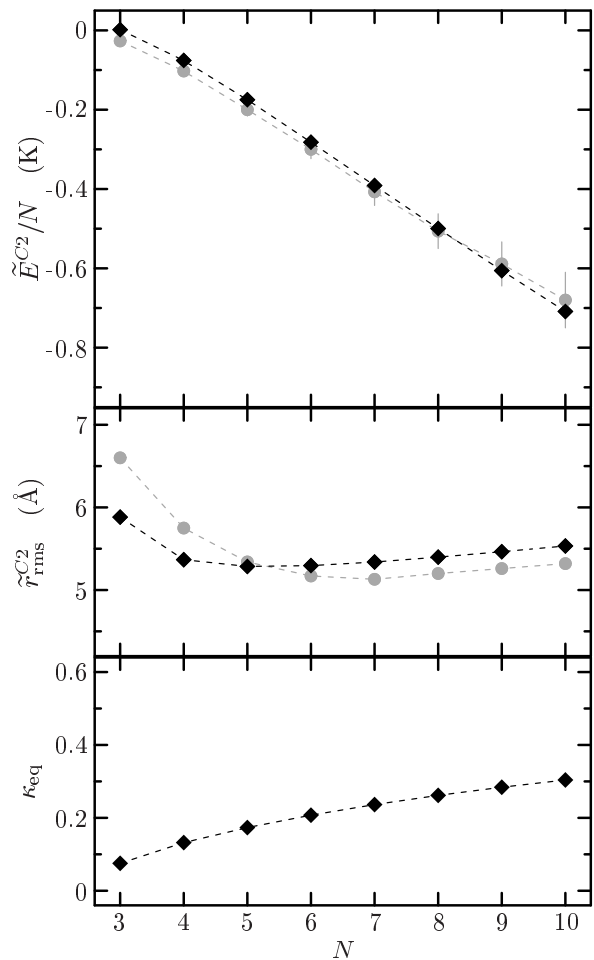}
  \caption{Energy per particle (upper panel), rms-Radius (middle panel),
    and smallness parameter $\kappa_{\equi}$ (lower panel) for small
    ${}^4$He droplets in two-body approximation with the
    density-dependent correlation function fixed in the homogeneous
    liquid. The gray circles show results of a variational Monte Carlo
    calculation \cite{ScSc65}. The lines should guide the eye.  }
  \label{fig:drop_2bdd}
\end{figure}

The results of the energy minimization with respect to the width
parameter $a$ are summarized in Figure \ref{fig:drop_2bdd}. The energy
expectation values are in full agreement with the VMC calculation
\cite{ScSc65} for all particle numbers. The overbinding observed with
a density-independent correlation function is completely compensated
by the density-dependence. The correlated rms-radii of droplets with
$N\ge6$ are consistently larger by 0.2\AA\ than the VMC
result. Unfortunately the authors \cite{ScSc65} give no estimate for
the statistical or systematic errors for their procedure to obtain the
rms-radii. For the smallest droplets our result for the rms-radii are
below the VMC results. This may be caused by the restriction to
Gaussian trial states \eqref{eq:drop_2b_wavefunc_sep}, which are not
able to describe the long-range exponential tail present in these
extremely weakly bound systems. At the same time the insufficient
trial state causes a slight underbinding of the smallest droplets.

For completeness Figure \ref{fig:drop_2bdd} also shows the smallness
parameter $\kappa_{\equi} = \rho_{\equi} V_C$. Since both, the
correlation volume $V_C$ and the density of the droplets are reduced
by the density-dependent correlator the product stays small. In any
case $\kappa_{\equi}$ has lost its meaning as a measure for three-body
correlations as the density-dependence of the correlator already
includes many-body correlations.

Table \ref{tab:drop} summarizes the results of the energy minimization
with two-body optimized and density-dependent correlator for droplet
sizes up to $N=70$. In addition to the correlated energy and the
rms-radius the individual terms of the correlated Hamiltonian are
shown. The attractive correlated potential and the repulsive local
part of the correlated kinetic energy are the major contributions and
show a large cancellation. The intrinsic kinetic energy and the
effective mass correction give rather small contributions.

\section{Summary and Conclusions}

We presented a simple and elegant approach to describe strong
short-range correlations in a many-body system that result from the
strong repulsive core of the two-body interaction. These correlations
induce high-momentum components into the relative wave function that
cannot be treated within a low-momentum model space. The correlation
operator $\CO$ generates the short-range correlations by a unitary
transformation in the relative coordinate of each pair of
particles. It shifts the particles out of the repulsive core of the
two-body interaction.

The correlation operator is then used to define correlated observables,
most notably the correlated Hamiltonian. The unitary transformation
yields a tamed effective two-body interaction, which consists of a purely
attractive local part and a weak momentum dependence. In addition
three-body and higher order terms are generated. For dilute systems,
i.e., if the range of the correlations is small compared to the average
distance between the particles, it is sufficient to include the terms
up to two-body order. In high-density systems the higher order
terms can be included explicitly or simulated by a density-dependent
correlation function. 

In order to study the applicability of the method in many-body
systems, where short-range correlations play a dominant role, we
investigated the ground state properties of homogeneous ${}^4$He
liquids and small droplets. The correlation function is fixed by
energy minimization in the two-body problem with a constant
uncorrelated wave function. Alternatively the correlation function
can be determined by mapping an uncorrelated wave function onto the
exact energy $E=0$ two-body wave function.  

We use this correlation function to calculate the energy of the
homogeneous ${}^4$He liquid as function of density. The uncorrelated
many-body state is a direct product of constant one-body states, thus
the influence of the interaction on the state has to be described by
the correlation operator alone. Already in two-body approximation the
strong short-range repulsion of the Lennard-Jones potential is tamed
completely and we obtain a bound system. However, the two-body
approximation does not generate saturation, i.e., the energy per
particle decreases proportional to density. To simulate the effect of
many-body correlations we employ a density-dependent correlation
function. The one parameter introduced by the density dependent
scaling of the correlation function is fixed by adjusting the
correlated energy for one value of the density to the result of a GFMC
calculation. The resulting equation of state shows a minimum and is in
very good agreement with the predictions of the GFMC calculation over
the whole density range.

In a second step we use the density-independent as well as the
density-dependent correlation function to perform an \emph{ab initio}
calculation of the ground state structure of small ${}^4$He droplets
in a simple variational framework. It turns out that the two-body
approximation with the density-independent correlation function gives
a good description for small droplets with particle numbers
$N\le6$. For larger droplets an overbinding compared to a quasi
exact variational Monte Carlo calculation appears. Due to the higher
density in these droplets many-body correlations become relevant. If
we use the density-dependent correlation function with one parameter
adjusted in the homogeneous liquid then the energies and rms-radii are
in very good agreement with the exact calculation for all particle
numbers under consideration.

We conclude that the Unitary Correlation Operator Method is a powerful
and universal tool to handle short-range correlations in the many-body
system. For sufficiently low densities one can use the two-body
approximation to construct the correlated Hamiltonian as well as other
correlated observables in closed form. The resulting effective
two-body interaction can be used for an ab initio description of the
many-body system based on a low-momentum model space.  A gross measure
for the validity of the two-body approximation is the smallness
parameter $\kappa$ \eqref{eq:ucom_smallnesspara}, which should be
below a phenomenological limit of $0.3$.

For higher densities one can utilize density-dependent correlation
functions to describe many-body correlations effectively within the
two-body approximation. The additional parameter in the
density-dependence can be fixed by one measured data point or by
comparison with an exact calculation.  Nevertheless, the resulting
density-dependent correlator can then be used for a broad range of
applications, e.g., \emph{ab initio} calculations of correlated
observables like energies, densities or momentum distributions of
different systems without further adjustments. The great advantage of
the Unitary Correlation Operator Method with density-dependent
correlation functions is that besides its transparent physics it
requires only minimal computational power compared to the huge
computational resources needed for Monte-Carlo-type calculations. A
set of \emph{Mathematica} notebooks that illustrate the application of
the Unitary Correlation Operator Method to $^4$He liquids can be
found at the URL: \texttt{http://theory.gsi.de/\~{}rroth/math/}.



\end{document}